\documentclass[aps,prl,twocolumn,longbibliography,floatfix]{revtex4-1}

\usepackage[T1]{fontenc}
\usepackage{newtxtext}
\usepackage{newtxmath}

\usepackage{amsmath}
\usepackage{mathtools}
\usepackage{amsfonts}
\usepackage{bm}

\usepackage{color}
\usepackage{xcolor}
\usepackage[colorlinks,
            citecolor=blue,
            linkcolor=red,
            urlcolor=blue]{hyperref}

\usepackage{graphicx}
\usepackage[all]{hypcap} 

\begin{document}

\title{Spatiotemporal Crossover between Low- and High-Temperature Dynamical Regimes\\ in the Quantum Heisenberg Magnet}

\author{Maxime Dupont}
\affiliation{Department of Physics, University of California, Berkeley, California 94720, USA}
\affiliation{Materials Sciences Division, Lawrence Berkeley National Laboratory, Berkeley, California 94720, USA}

\author{Nicholas E. Sherman}
\affiliation{Department of Physics, University of California, Berkeley, California 94720, USA}
\affiliation{Materials Sciences Division, Lawrence Berkeley National Laboratory, Berkeley, California 94720, USA}

\author{Joel E. Moore}
\affiliation{Department of Physics, University of California, Berkeley, California 94720, USA}
\affiliation{Materials Sciences Division, Lawrence Berkeley National Laboratory, Berkeley, California 94720, USA}

\begin{abstract}
    The stranglehold of low temperatures on fascinating quantum phenomena in one-dimensional quantum magnets has been challenged recently by the discovery of anomalous spin transport at high temperatures. Whereas both regimes have been investigated separately, no study has attempted to reconcile them. For instance, the paradigmatic quantum Heisenberg spin-$1/2$ chain falls at low temperature within the Tomonaga-Luttinger liquid framework, while its high-temperature dynamics is superdiffusive and relates to the Kardar-Parisi-Zhang universality class in $1+1$ dimensions. This Letter aims at reconciling the two regimes. Building on large-scale matrix product state simulations, we find that they are connected by a temperature-dependent spatiotemporal crossover. As the temperature $T$ is reduced, we show that the onset of superdiffusion takes place at longer length and timescales $\propto 1/T$. This prediction has direct consequences for experiments including nuclear magnetic resonance: it is consistent with earlier measurements on the nearly ideal Heisenberg $S=1/2$ chain compound Sr$_2$CuO$_3$, yet calls for new and dedicated experiments.
\end{abstract}

\maketitle

\textit{Introduction.---} At low temperatures, reduced spatial dimensionality greatly enhances quantum fluctuations in physical systems, giving rise to exotic properties. In that regard, one-dimensional (1D) quantum many-body systems have always been influential and generically fall into two classes~\cite{PhysRevLett.50.1153,giamarchi2003quantum}: on the one hand, gapless low-energy excitations described in the framework of Tomonaga-Luttinger liquid (TLL), and on the other, a gapped behavior. Theoretical predictions have been intensively checked by experiments in various contexts, ranging from ultracold atom setups to quantum magnets~\cite{giamarchi2013,wierschem2014}.

At energy $\hbar\omega\ll k_\mathrm{B}T$, the physics is usually thought of in terms of thermal rather than quantum effects. This regime had not been thought to hold phenomena as compelling as its low-temperature counterpart until very recently. Indeed, recent theoretical progress suggests that the equilibrium and out-of-equilibrium dynamics of some 1D quantum systems can exhibit peculiar behaviors and contain information about the intrinsic quantum features, even at very high temperatures~\cite{bertini2016,castroalvaredo2016,bulchandani2018}.

While such many-particle systems are governed at the microscopic level by the Schr\"odinger equation, they display in the long-time and long-wavelength limits an emergent coarse-grained hydrodynamic behavior. An analogy can be made with classical fluid dynamics: one does not describe individual particles with Newton's laws of motion but relies instead on phenomenological continuous differential equations, ideally more amenable. The derivation of hydrodynamic equations is based essentially on continuity equations of conserved quantities (e.g., mass, energy, etc.), assuming local equilibrium~\cite{landau1987}.

Quantum systems also possess conservation laws, and depending on those, one expects the emergence of different kinds of coarse-grained hydrodynamic descriptions. Singularly in 1D, a class of quantum systems---known as integrable---has an infinite set of nontrivial conserved quantities that can lead to anomalous dynamical behaviors~\cite{zotos1997,sirker2006,sirker2011,prosen2011,znidaric2011,karrasch2013,ilievski2015,bertini2016,castroalvaredo2016,bulchandani2018,denardis2018,gopalakrishnan2018,denardis2019scipost,agrawal2019,ljubotina2019,denardis2019,gopalakrishnan2019,agrawal2019,Gopalakrishnan16250,dupont2020,denardis2020,denardis2020b,friedman2020,agrawal2020,ilievski2020,bulchandani2020,bertini2021,lopezpiqueres2021,denardis2021,scheie2021}.

Integrable systems are typically described by very fine-tuned models but some of them can be reliably realized in the lab (e.g., the Lieb-Liniger model representing a gas of one-dimensional bosons with contact repulsion~\cite{PhysRev.130.1605,PhysRev.130.1616}) and found with high fidelity in nature (e.g., the spin-$1/2$ Heisenberg chain of magnetic moments coupled by a nearest-neighbor exchange interaction~\cite{giamarchi2003quantum}). In that context, some of the theoretical predictions have been successfully tested on 1D cloud of trapped ${}^{87}$Rb~\cite{schemmer2019,malvania2020} and ${}^{7}$Li~\cite{jepsen2020} atoms for out-of-equilibrium dynamics and by neutron scattering on the quantum magnet KCuF$_3$ at thermal equilibrium~\cite{scheie2021}.

In the case of quantum magnets, it has been numerically conjectured, based on microscopic simulations, that in the limit of infinite temperature, the spin dynamics of the $S=1/2$ Heisenberg chain is anomalous and belongs to the Kardar-Parisi-Zhang (KPZ) universality class in $1+1$ dimensions~\cite{kardar1986,ljubotina2019}. It is characterized by a dynamical exponent $z=3/2$, controlling the length-time scaling of the dynamical properties. This exponent has been recently observed in the high-temperature neutron spectrum of KCuF$_3$~\cite{scheie2021}, which is directly proportional to the dynamical structure factor, probing spin-spin correlations.

\begin{figure}[!t]
    \includegraphics[width=1.0\columnwidth]{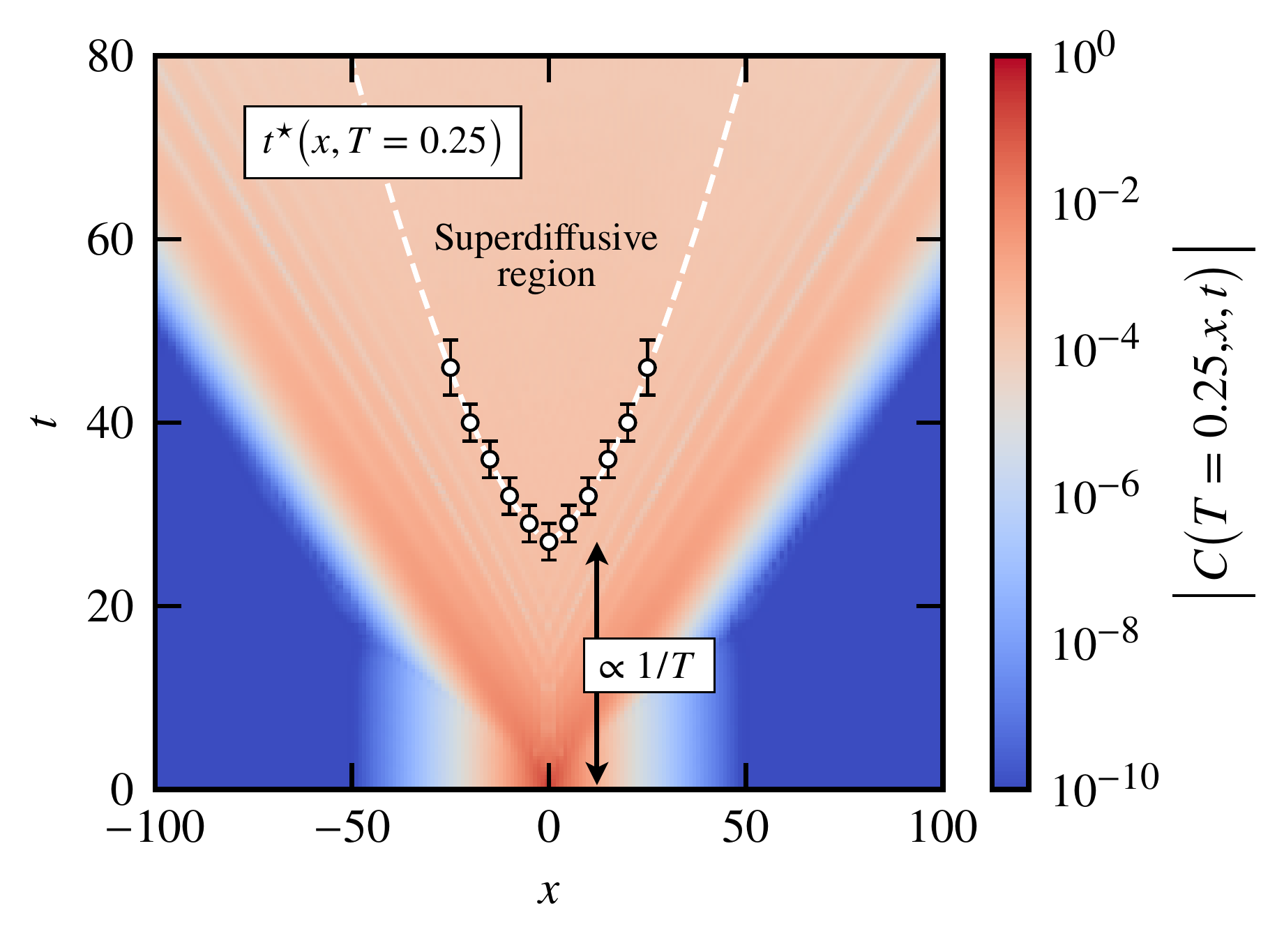} 
    \caption{Log-scale intensity plot of the Euclidean norm of the spin-spin correlation~\eqref{eq:corr_def} at $T=0.25$. Simulation obtained for $L=256$ with $\chi=1024$. The goal of this Letter is to determine and study the superdiffusive region delimited by the spatiotemporal crossover $t^\star$ of Eq.~\eqref{eq:corr_hydro} versus the temperature (white circles and dashed white line). As the temperature is decreased, we find that the superdiffusive region is shifted vertically to longer and longer times by a factor $\propto 1/T$, and eventually disappears at exactly zero temperature.}
    \label{fig:colormap}
\end{figure}

Here, we seek to reconcile the low-temperature physics of the $S=1/2$ Heisenberg chain, falling within the gapless TLL category, with the recently found infinite-temperature KPZ hydrodynamics. Whereas both regimes have been studied independently, no work has attempted to bring them together. In this Letter, we precisely define the long-time and long-wavelength limits for the emergence of anomalous dynamics versus the temperature. We find that these limits define a spatiotemporal crossover beyond which hydrodynamics take place. As the temperature is lowered, the crossover is pushed toward infinity and eventually disappears at exactly zero temperature, see Fig.~\ref{fig:colormap}. This scenario allows one to recover the well-known zero temperature results where KPZ hydrodynamics is absent. Moreover, because experimental dynamical condensed matter probes such as neutron scattering or nuclear magnetic resonance (NMR) work for all practical purposes at a finite frequency and finite temperatures, it is paramount to better understand and quantitatively define the theoretical limits. We discuss the implication of our results for experiments and confront our findings to earlier high-temperature NMR experiments on the nearly ideal Heisenberg spin-$1/2$ compound Sr$_2$CuO$_3$~\cite{PhysRevLett.87.247202}.

\textit{Model and method.---} The 1D spin-$1/2$ Heisenberg model is described by the lattice Hamiltonian,
\begin{equation}
    \hat{\mathcal{H}}=J\sum\nolimits_j\hat{\boldsymbol{S}}_j\cdot\hat{\boldsymbol{S}}_{j+1},
    \label{eq:hamiltonian}
\end{equation}
with $\hat{\boldsymbol{S}}_j=(\hat{S}^x_j, \hat{S}^y_j, \hat{S}^z_j)$ and $J>0$ the nearest-neighbor antiferromagnetic exchange. To investigate the thermal equilibrium spin dynamics, we consider the time-dependent spin-spin correlation function
\begin{equation}
    C\bigl(T,x,t\bigr)=\mathrm{tr}\bigl[\hat{\boldsymbol{S}}_x\bigl(t\bigr)\cdot\hat{\boldsymbol{S}}_{0}\bigl(0\bigr)\,\hat{\rho}_T\bigr]~\in\mathbb{C},
    \label{eq:corr_def}
\end{equation}
with $\hat{\rho}_T=\mathrm{e}^{-\hat{\mathcal{H}}/k_\mathrm{B}T}/\mathrm{tr}(\mathrm{e}^{-\hat{\mathcal{H}}/k_\mathrm{B}T})$ as the thermal density matrix of the system at temperature $T$ and $\hat{\boldsymbol{S}}_j\bigl(t\bigr)=\mathrm{e}^{i\hat{\mathcal{H}}t/\hbar}\hat{\boldsymbol{S}}_j\mathrm{e}^{-i\hat{\mathcal{H}}t/\hbar}$ as the time-dependent spin operator in the Heisenberg picture. We set $J=k_\mathrm{B}=\hbar=1$ in the following. We compute the correlation function~\eqref{eq:corr_def} based on a numerical matrix product state (MPS) approach~\cite{schollwock2011,itensor}, where we represent the mixed state as a pure state in an enlarged Hilbert space~\cite{PhysRevLett.93.207204,PhysRevLett.93.207205}. We use the time-evolving block decimation algorithm~\cite{PhysRevLett.93.040502} along with a fourth-order Trotter decomposition~\cite{hatano2005} to handle the exponential operators~\cite{trotter}. To ensure convergence of the numerical data, we study in the Supplemental Material the effect of the bond dimension $\chi$ of the MPS, which is the control parameter of the simulations (larger is better, but computationally more expensive)~\cite{supplemental}.

At fixed distance $x$ and temperature $T$, the hydrodynamics regime is characterized by an algebraic decay of the Euclidean norm of the spin-spin correlation~\eqref{eq:corr_def} function at long time,
\begin{equation}
    \bigl\vert C\bigl(T,x,t\bigr)\bigr\vert\propto t^{-1/z}\quad\mathrm{for}~~t\gtrsim t^\star\bigl(x, T\bigr),
    \label{eq:corr_hydro}
\end{equation}
with $z$ as the dynamical exponent. The long-time limit is denoted by the crossover time $t^\star$, which we aim to identify, see Fig.~\ref{fig:colormap}. Depending on the microscopic model, three values for the exponent $z$ have been reported for 1D quantum magnets: $z=3/2$ corresponding to superdiffusion, $z=1$ for ballistic, and $z=2$ for diffusion~\cite{dupont2020,denardis2020}. Superdiffusion is expected for the isotropic spin-$1/2$ Heisenberg model of Eq.~\eqref{eq:hamiltonian}.

\begin{figure}[!t]
    \includegraphics[width=1.0\columnwidth]{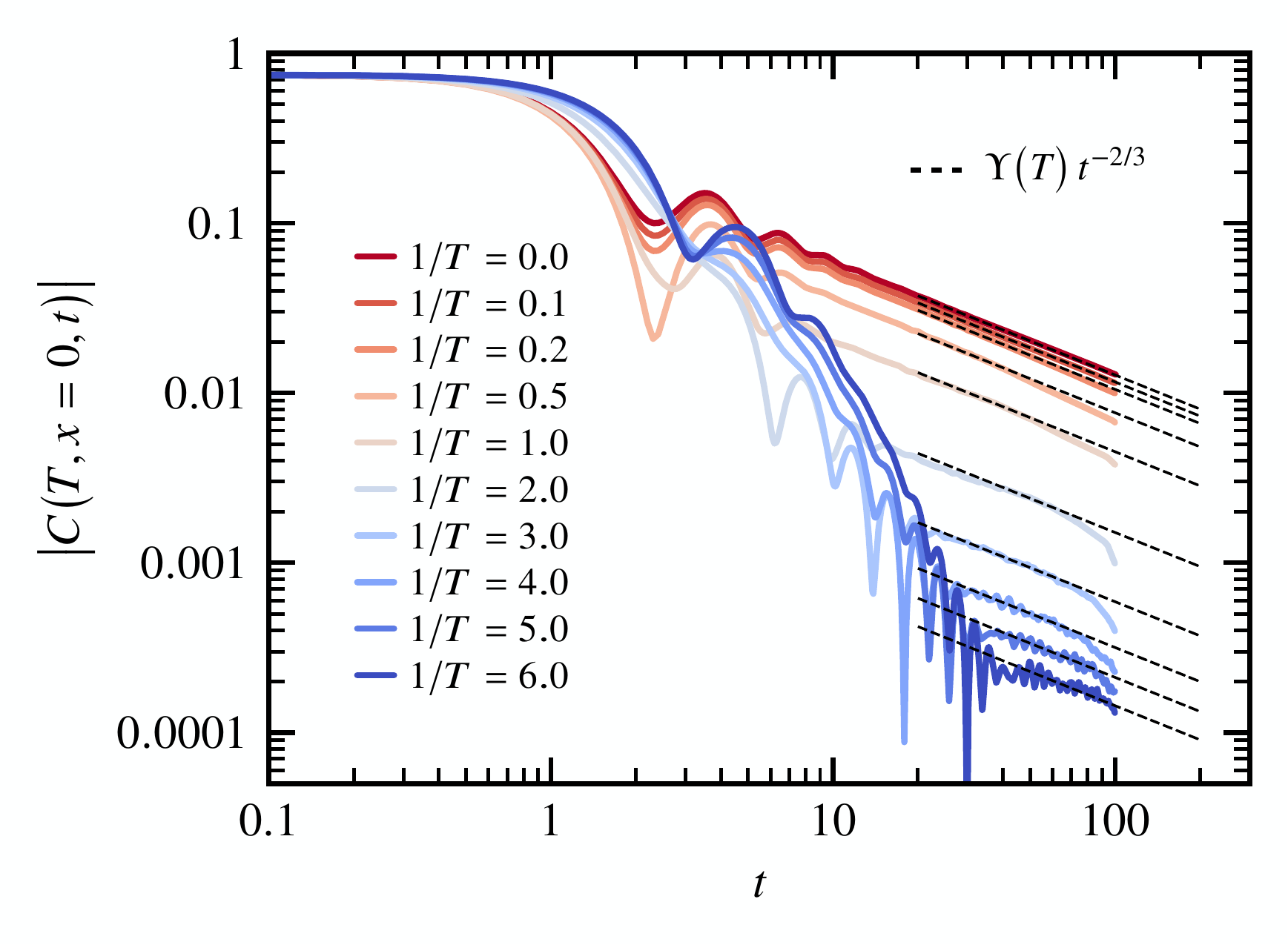} 
    \caption{Time dependence of the norm of the spin-spin correlation~\eqref{eq:corr_def} at $x=0$ for various temperatures $T$. Simulations obtained for $L=256$ with $\chi=1024$. At long time, it displays an algebraic decay with time, according to Eq.~\eqref{eq:corr_hydro}. It is well fitted by the form $\Upsilon(T)\,t^{-2/3}$ with $\Upsilon(T)$, a temperature-dependent prefactor decreasing with the temperature reported in Fig.~\ref{fig:temperature_dependence}(b). The deviation from the genuine power law at long time is the result of the bond dimension being too small~\cite{supplemental}.}
    \label{fig:autocorrelation}
\end{figure}

\textit{Autocorrelation.---} We first consider the autocorrelation function ($x=0$) versus time for different temperatures, as plotted in Fig.~\ref{fig:autocorrelation}. Two regimes are clearly visible, delimited by the crossover time $t^\star(x=0, T)$~\cite{supplemental}. Beyond the crossover time and for all temperatures, one finds the expected power-law decay $\propto t^{-2/3}$ of superdiffusive hydrodynamics. Note that the rapid change of slope from the genuine power-law, at the longest times displayed, is the result of the bond dimension being too small and not a physical effect~\cite{supplemental}.

With high-temperature physics beyond $t^\star$, one can suspect low-temperature features at shorter times. For instance, the oscillating behavior observed in the norm of the autocorrelation is reminiscent of a change of sign in the real and imaginary part~\cite{supplemental}, signaling antiferromagnetic correlations as the temperature is lowered. The long-time asymptotic of $C(T=0,x=0,t)$ has been studied at exactly zero temperature~\cite{pereira2008,pereira2012}. It is composed by several power-law decaying contributions with the slowest one being $\propto t^{-1}$ (up to logarithmic corrections inherent to the isotropic spin-$1/2$ Heisenberg antiferromagnet~\cite{affleck1989,nomura1991,eggert1994,takigawa1997,affleck1998,barzykin2000,barzykin2001,dupont2016,supplemental}). We cannot identify this regime in Fig.~\ref{fig:autocorrelation}, which we attribute to insufficiently low temperatures; see the Supplemental Material for additional data~\cite{supplemental}.

\begin{figure}[!t]
    \includegraphics[width=1.0\columnwidth]{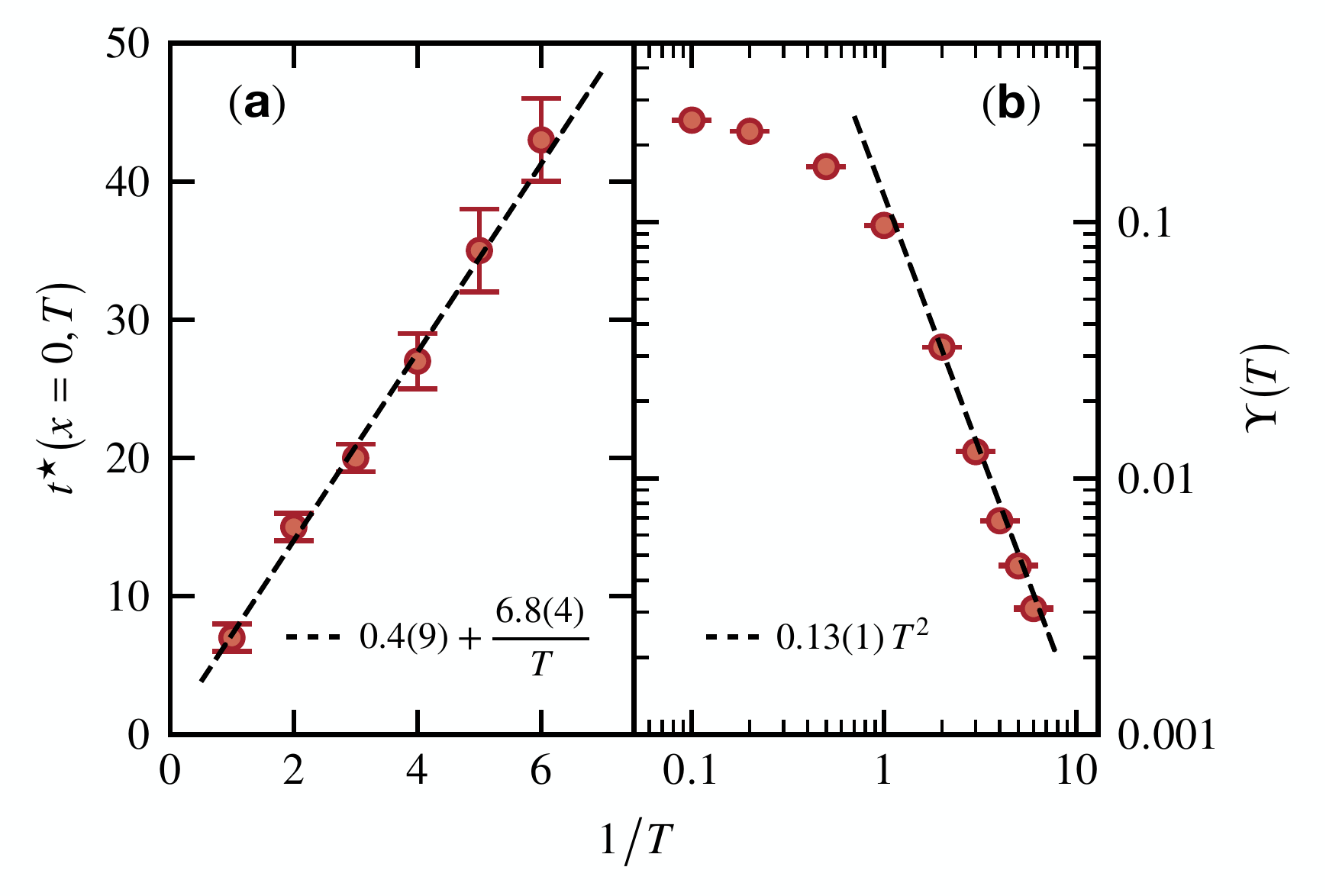} 
    \caption{The data points are extracted from Fig.~\ref{fig:autocorrelation}. (a) Temperature dependence of the crossover timescale $t^\star\bigl(x=0,T\bigr)$ beyond which the algebraic decay $\propto t^{-2/3}$ for superdiffusive hydrodynamics emerges, see Eq.~\eqref{eq:corr_hydro}. It shows a linear dependence with the inverse temperature (dashed line). (b) Temperature dependence of the prefactor $\Upsilon(T)$ of the algebraic decay $\propto t^{-2/3}$ for superdiffusive hydrodynamics. At low temperatures $T\lesssim 1$, it follows a quadratic dependence $\propto T^2$ (dashed line).}
    \label{fig:temperature_dependence}
\end{figure}

We now turn our attention to the temperature dependence of the crossover time $t^\star(x=0, T)$. It is plotted in Fig.~\ref{fig:temperature_dependence}(a) versus the inverse temperature and shows a linear dependence. It can be understood as follows. It is well known that a finite temperature induces a thermal correlation length $\xi$ which diverges as $T\to 0$ as $\propto u/T$ (up to logarithmic corrections~\cite{nomura1991,supplemental}) with $u$ the velocity of low-energy excitations in the spin-$1/2$ chain. Moreover, the dynamical correlation function~\eqref{eq:corr_def} can also be thought of as measuring the spreading of a spin excitation. In this picture, the system behaves like a TLL for $t\lesssim\xi/u$, which can be identified as the crossover time $t^\star(x=0, T)\propto 1/T$. Hence, the onset of superdiffusive hydrodynamics simply takes place as the low-energy physics gets suppressed by the finite temperature. It is only at zero temperature that the system is strictly critical and thus does not display any sign of anomalous high-energy dynamics. In addition to the linear dependence with $\propto 1/T$, there is an $O(1)$ constant in Fig.~\ref{fig:temperature_dependence}(a) that coincides with the very short-time dynamics where $\vert C(T,x=0,t\simeq 0)\vert\simeq 0.75$.

At infinite temperature, it has been established that the dynamics belong to the $1+1$ KPZ universality class~\cite{kardar1986,ljubotina2019}, as it shows the same scaling laws as appear in the KPZ equation itself: $\partial_th=\frac{1}{2}\lambda\bigl(\partial_x h\bigr)^2+\nu\partial_x^2h+\sqrt{\sigma}\eta$ with $h\equiv h(x,t)$, $\eta\equiv\eta(x,t)$ a normalized Gaussian white noise, and $\lambda$, $\nu$, and $\sigma$ parameters. It is a Langevin equation, with no quantum roots---and which makes the observation of its physics in a quantum magnet rather puzzling. In the right limits, the noise-averaged slope correlations behave as~\cite{spohn2014,spohn2016}
\begin{equation}
    C_\mathrm{KPZ}\bigl(x,t\bigr)\simeq \chi_\mathrm{s}\bigl(\lambda_\mathrm{KPZ}t\bigr)^{-2/3}f_\mathrm{KPZ}\left[x\bigl(\lambda_\mathrm{KPZ}t\bigr)^{-2/3}\right],
    \label{eq:kpz_scaling}
\end{equation}
with $\chi_\mathrm{s}=\sigma/2\nu$ as the static spin susceptibility~\cite{supplemental}, $\lambda_\mathrm{KPZ}=\sqrt{2}\lambda$, and $f_\mathrm{KPZ}$ as the KPZ scaling function~\cite{prahofer2004}. The numerical observation of the scaling~\eqref{eq:kpz_scaling} for the Heisenberg spin chain through the spin-spin correlation~\eqref{eq:corr_def} served as a conjecture regarding the nature of its dynamics~\cite{ljubotina2019}. A theoretical scenario for how KPZ hydrodynamics emerges in the Heisenberg chain has been advanced~\cite{bulchandani2020}. A relation between the parameters of the KPZ equation with those of the microscopic quantum model has been proposed~\cite{denardis2020b}. Here, by identifying the prefactor of $C_\mathrm{KPZ}(x=0,t)$ in Eq.~\eqref{eq:kpz_scaling} with the prefactor $\Upsilon(T)$ of the power-law decay $\propto t^{-2/3}$ shown in Fig.~\ref{fig:temperature_dependence}(b), we are able to report on the temperature dependence of the parameters. The high-temperature data points are compatible with Ref.~\onlinecite{denardis2020b}. In addition, for $T\lesssim 1$, we find that $\Upsilon(T)=0.13(1)T^2$, and therefore that $\chi_\mathrm{s}\lambda_\mathrm{KPZ}^{-2/3}f_\mathrm{KPZ}(0)\propto T^2$. We argue in the following that this behavior is compatible with earlier NMR experiments on Sr$_2$CuO$_3$~\cite{PhysRevLett.87.247202,supplemental}.

The definition of the crossover time $t^\star$ in Eq.~\eqref{eq:corr_hydro} for the onset of superdiffusion is related to the power-law dependence $\propto t^{-2/3}$ and not $f_\mathrm{KPZ}$ of Eq.~\eqref{eq:kpz_scaling}. It is well known that unambiguously identifying the scaling function from microscopic simulations with $f_\mathrm{KPZ}$ requires great numerical precision and long-time data for all distances $x$~\cite{ljubotina2019}. This is beyond the capability of our simulations at low temperatures. Instead, we consider the spatial dependence of $t^\star$ for $\vert{x}\vert>0$.

\begin{figure}[!t]
    \includegraphics[width=1.0\columnwidth]{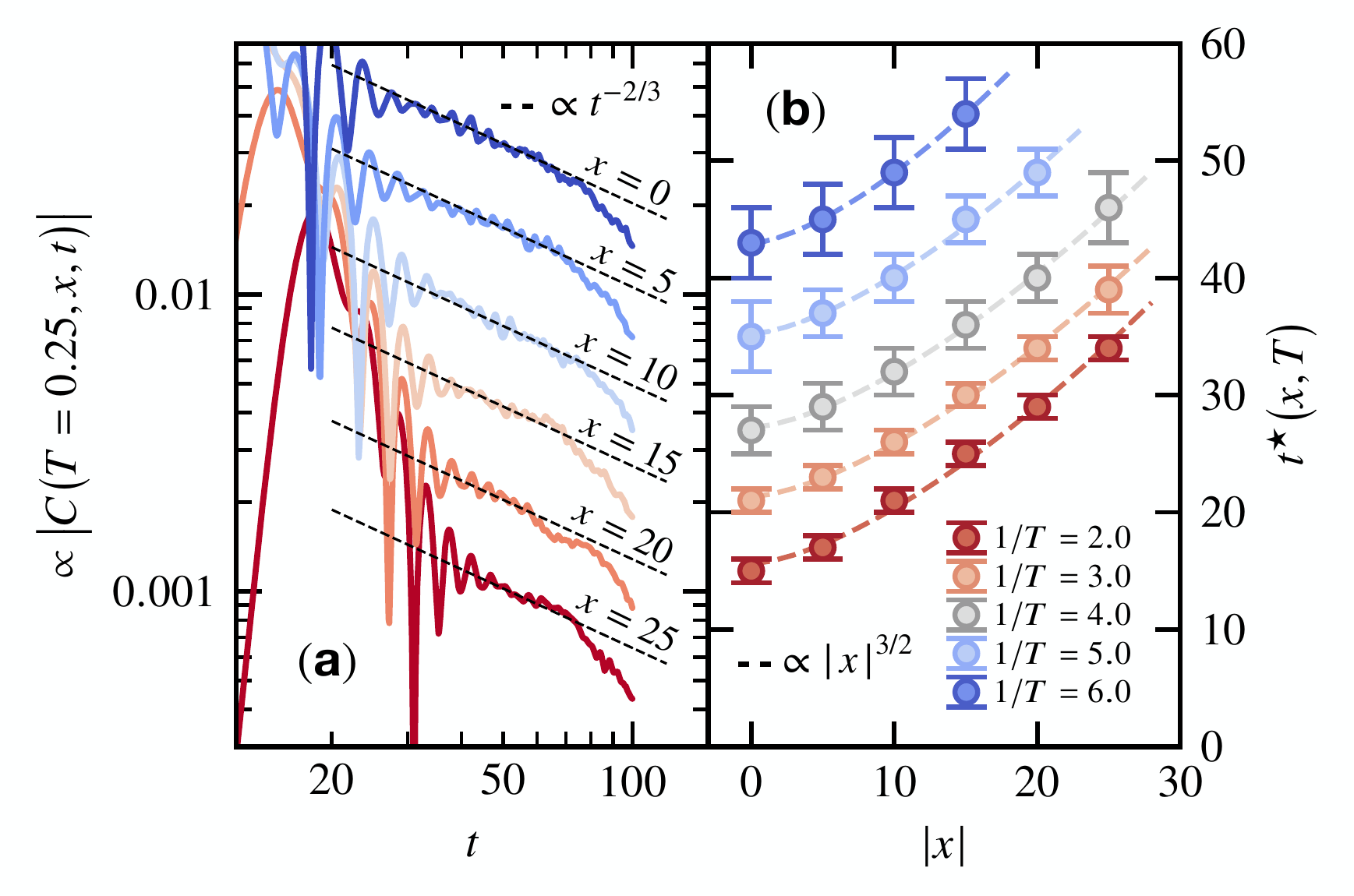} 
    \caption{(a) Time dependence of the norm of the spin-spin correlation~\eqref{eq:corr_def} at $T=0.25$ for various distances $x$. Simulations obtained for $L=256$ with $\chi=1024$. The curves have been shifted vertically for visibility. At long time, it displays an algebraic decay with time, according to Eq.~\eqref{eq:corr_hydro}, well fitted by the form $\propto t^{-2/3}$. The deviation from the genuine power law at long time is the result of the bond dimension being too small~\cite{supplemental}. (b) Spatial dependence of the crossover time $t^\star(x,T)$ beyond which the algebraic decay $\propto t^{-2/3}$ for superdiffusive hydrodynamics emerges, see Eq.~\eqref{eq:corr_hydro}. The dashed lines are fits of the form $A+B\vert{x}\vert^{3/2}$ with $A\equiv t^\star(0,T)$ and $B=0.17(3)$ found to be temperature independent~\cite{supplemental}.}
    \label{fig:spatiotemporal_crossover_data}
\end{figure}

\textit{Spatiotemporal crossover.---} The time-dependent spin-spin correlation function~\eqref{eq:corr_def} is associated with a light-cone structure and we therefore expect $t^\star(x,T)$ to be an increasing function with the distance $\vert{x}\vert$. It is verified in Fig.~\ref{fig:spatiotemporal_crossover_data}(a) where we plot its time dependence at fixed temperature ($T=0.25$). As $\vert{x}\vert$ increases, the onset of superdiffusion takes place at longer and longer times, and we display the crossover timescale in Fig.~\ref{fig:spatiotemporal_crossover_data}(b) for different temperatures. Because we can only reliably estimate it for $\vert{x}\vert\lesssim 30$, it is difficult to draw a definite conclusion on its scaling. Nevertheless it is compatible with a superdiffusive length-time scaling of the form,
\begin{equation}
    t^\star\bigl(x,T\bigr) = 0.4(9) ~+~ \frac{6.8(4)}{T} ~+~ 0.17(3)\,\bigl\vert\,{x}\,\bigr\vert^{3/2},
    \label{eq:crossover_form}
\end{equation}
with the first two terms obtained from the $t^\star(x=0,T)$ data, see Fig.~\ref{fig:temperature_dependence}(a). The prefactor of $\vert{x}\vert^{3/2}$ is found independent of the temperature~\cite{supplemental}. The reported numerical parameters are obtained by least-square fitting. The spatiotemporal crossover time~\eqref{eq:crossover_form} is plotted on top of the norm of the spin-spin correlation in Fig.~\ref{fig:colormap} for $T=0.25$. Note that based on this picture, we expect logarithmic corrections for the temperature dependence, but they are not detectable from our simulations~\cite{logcorrections}.

\textit{Experimental consequences.---} Although we have focused on the norm of the spin-spin correlation~\eqref{eq:corr_def}, we find that $\vert\mathfrak{Im}\,C(T,x,t)\vert\ll\vert\mathfrak{Re}\,C(T,x,t)\vert$ for $t\gtrsim t^\star$, and that the superdiffusive power law $\propto t^{-2/3}$ only holds for the real part~\cite{supplemental}, which therefore hosts the relevant high-temperature physics. For instance, superdiffusion was observed in KCuF$_3$ by neutron scattering in the limit of small momentum and vanishing frequency~\cite{scheie2021}, which probes the Fourier transform to momentum and frequency spaces of $C(T,x,t)$.

Another promising experimental technique for investigating high-temperature hydrodynamics is NMR, which has been successfully used to characterize the low-temperature TLL regime in numerous spin compounds~\cite{klanjsek2008,bouillot2011,jeong2013,jeong2016,dupont2016,coira2016,berthier2017,dupont2018,horvatic2020}. Nuclear spins are polarized via a static magnetic field (ideally weak) and then perturbed by an electromagnetic pulse of frequency $\omega_0$, chosen to target specific nuclei as per the Zeeman splitting. Following the perturbation, the nuclear spins relax over time with an energy transfer to the electrons. When the nuclear and electronic spins belong to the same atom, the relaxation rate is related to the autocorrelation function, $1/T_1\sim\int_0^{1/\omega_0}\mathfrak{Re}\,C(T,x=0,t)\,\mathrm{d}t$~\cite{abragam1961,horvatic2002,slichter2013}. With $\omega_0$ of the order of a few mK, it usually leads to a frequency-independent $1/T_1$ as long as the correlation decays quickly enough. Here, the hydrodynamics regime should lead instead to $1/T_1\propto\omega_0^{1/z-1}$ and give access to $z$ in the right frequency regime. According to Eq.~\eqref{eq:crossover_form}, the corresponding crossover frequency scale $\omega^\star\sim 1/t^\star$ goes as $\propto T$, and superdiffusion will be visible if $\omega_0\ll\omega^\star\sim T$. Considering the experimental range of $\omega_0$, this condition is fulfilled even at low temperatures, where measurements are often less noisy and less subject to spoiling effects such as phonons.

Thus, the existence of a finite spatiotemporal crossover $t^\star(x,T)$ in the form of Eq.~\eqref{eq:crossover_form} confirms that superdiffusive hydrodynamics is within the experimentally relevant window of parameters with respect to temperatures, time and length scales for quantities involving $\mathfrak{Re}\,C(T,x,t)$.

In fact, a power-law behavior of the form $1/T_1\propto\omega_0^{-\alpha}$ has been reported in the nearly ideal spin-$1/2$ Heisenberg antiferromagnet Sr$_2$CuO$_3$ ($J\simeq 2200$ K) at $T=295$ K a couple of decades ago~\cite{PhysRevLett.87.247202}. NMR was performed on the ${}^{17}$O, coupled symmetrically to the Cu$^{2+}$ carrying the relevant electronic spin, which filtered out the $q=\pm\pi$ contributions in the $1/T_1$ due to form factors, but not the long-wavelength modes $q=0$ holding hydrodynamics. Although the measurement accuracy was not sufficiently precise to extract the exponent $\alpha$, the results are compatible with $\alpha\approx 0.33$, which corresponds to $z=3/2$~\cite{supplemental}. In addition, the authors find that at fixed frequency, the NMR relaxation rate may be approximated by an empirical form $1/T_1T\approx a+bT$ for $T\ll J$ with $a$ and $b$ fitting constants. When dropping $a$, this is compatible with $\Upsilon(T)\propto T^2$ reported in Fig.~\ref{fig:temperature_dependence}(b)~\cite{supplemental}, which relates to the temperature dependence of the parameters of the KPZ equation.

Today's theoretical understanding of the dynamics of 1D quantum systems and our results call for new NMR experiments on spin chains at high temperatures. It would provide a complementary probe to neutron scattering~\cite{scheie2021} to access anomalous spin transport in quantum materials.

\textit{Conclusion.---} Building on large-scale MPS calculations, we reconciled the well-established low-temperature dynamics of the quantum Heisenberg spin-$1/2$ chain with the recently predicted high-temperature superdiffusive regime related to KPZ hydrodynamics. We have found that both coexist, and the transition from one to the other takes the form of a spatiotemporal crossover. The crossover is controlled by the temperature: as the temperature is lowered, the growing quantum correlations between degrees of freedom push the onset of superdiffusion to longer length and timescales as $\propto 1/T$. We also reported on the temperature dependence of the parameters of the KPZ equation, which should provide useful guidance in relating them to the microscopic parameters of the quantum model. We also showed that only the real part of the spin-spin correlations holds the superdiffusive hydrodynamics. Finally, we discussed the experimental consequences of our results for condensed matter probes. We motivated NMR experiments as a great way to measure spin transport in quantum materials and showed that earlier results are compatible with the current theoretical understanding yet calling for new experiments in quantum spin chains. Because NMR requires the use of a static magnetic field to polarize the nuclear spins, it would be insightful to study the effect of this perturbation on the dynamics of the $S=1/2$ Heisenberg chain studied in this Letter. We believe that it would induce another crossover from superdiffusion to ballistic dynamics, which needs to be characterized.

\begin{acknowledgments}
    We gratefully acknowledge G.E. Granroth, S.E. Nagler, A. Scheie, M.B. Stone, and D.A. Tennant for collaborations on related works. We acknowledge discussions with S. Brown. M.D. acknowledges discussions with J. De Nardis, S. Gopalakrishnan, and R. Vasseur. M.D. was supported by the U.S. Department of Energy, Office of Science, Office of Basic Energy Sciences, Materials Sciences and Engineering Division under Award No. DE-AC02-05-CH11231 through the Scientific Discovery through Advanced Computing (SciDAC) program (KC23DAC Topological and Correlated Matter via Tensor Networks and Quantum Monte Carlo). N.S. and J.E.M. were supported by the U.S. Department of Energy, Office of Science, Office of Basic Energy Sciences, Materials Sciences and Engineering Division under Award No. DE-AC02-05-CH11231 through the Theory Institute for Molecular Spectroscopy (TIMES). J.E.M. was also supported by a Simons Investigatorship. This research used the Lawrencium computational cluster resource provided by the IT Division at the Lawrence Berkeley National Laboratory (supported by the Director, Office of Science, Office of Basic Energy Sciences, of the U.S. Department of Energy under Award No. DE-AC02-05CH11231). This research also used resources of the National Energy Research Scientific Computing Center (NERSC), a U.S. Department of Energy Office of Science User Facility operated under Award No. DE-AC02-05CH11231.
\end{acknowledgments}

\bibliography{references}

\begin{thebibliography}{75}
\bibitem [{tro()}]{trotter}%
  \BibitemOpen
  \href@noop {} {}\bibinfo {note} {We use a Trotter step $\delta=0.1$ leading
  to a negligible discretization error $O\bigl(\delta^5\bigr)$.}\BibitemShut
  {Stop}%
\end{thebibliography}

\onecolumngrid
\setcounter{secnumdepth}{3}
\setcounter{figure}{0}
\setcounter{equation}{0}
\renewcommand\thefigure{S\arabic{figure}}
\renewcommand\theequation{S\arabic{equation}}
\newpage\clearpage

\setlength{\belowcaptionskip}{0pt}

\begin{center}
    \large\textbf{Supplemental Material for ``\textit{Spatiotemporal Crossover between\\ Low- and High-Temperature Dynamical Regimes in the Quantum Heisenberg Magnet''}}
\end{center}

\addvspace{5mm}
\begin{center}
    \begin{minipage}{0.8\textwidth}
        First, we revisit past NMR measurements for the nearly-ideal spin-$1/2$ Heisenberg chain Sr$_2$CuO$_3$. We show that these results, which were interpreted in the context of diffusion, are compatible with the current understanding of the high-temperature dynamics of the one-dimensional $S=1/2$ Heisenberg model, which is known to be superdiffusive. Second, we provide additional data to understand the convergence of the numerical simulation with respect to the control parameter (namely the bond dimension of the matrix product state $\chi$) and the system size $L$. Third, we show that the real part of the dynamical spin-spin correlation function dominates the imaginary part and that the real part hosts the characteristic power-law dependence $\propto t^{-2/3}$. Fourth, we present additional numerical results at exactly zero temperature to connect our low-temperature data to zero temperature dynamics. Fifth, we provide data on the temperature dependence of the correlation length $\xi$ of the spin-$1/2$ Heisenberg chain. Sixth, we plot the temperature dependence of the static spin susceptibility $\chi_\mathrm{s}$ of the spin-$1/2$ Heisenberg chain and discuss further the temperature dependence of the parameters of the KPZ equation. Seventh, we discuss the spatial dependence $\propto\vert{x}\vert^{3/2}$ of the crossover time $t^\star\bigl(x,T\bigr)$. Finally, we provide information on how the crossover time $t^\star\bigl(x,T\bigr)$ is extracted from the numerical simulations.
    \end{minipage}
\end{center}

\section{Revisiting experimental NMR data for Sr$_2$CuO$_3$}

\subsection{Characterizing anomalous spin transport}

\begin{figure}[!h]
    \includegraphics[width=0.5\columnwidth]{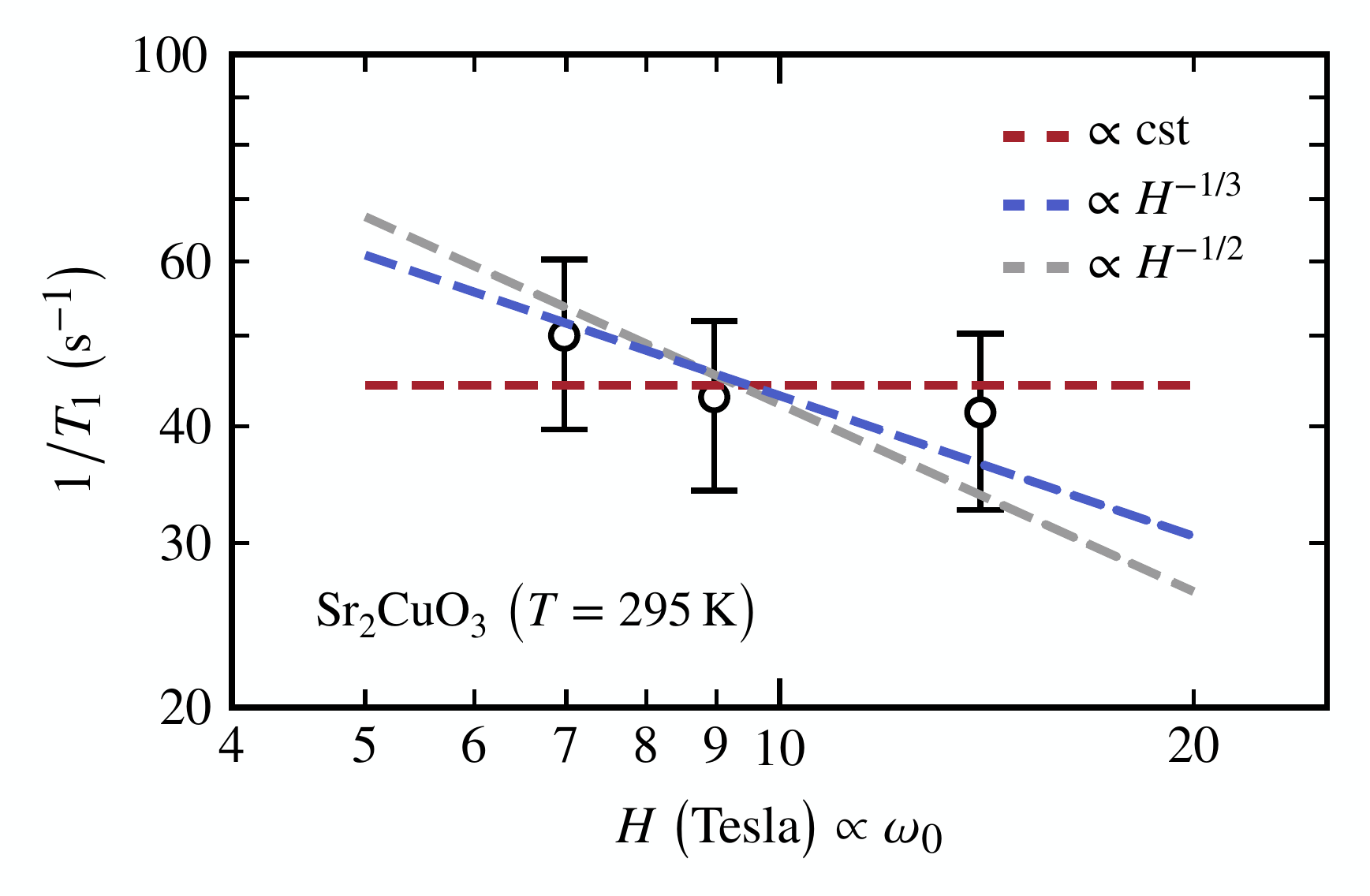} 
    \caption{The data reported on this figure is extracted from Fig.~3(d) of Ref.~\onlinecite{PhysRevLett.87.247202}. It corresponds to the NMR relaxation rate $1/T_1$ versus the strength of the applied external magnetic field $H$ for Sr$_2$CuO$_3$ at $T=295$ K (the exchange coupling is $J\simeq 2200$ K). The NMR was performed on the ${}^{17}$O nuclei, coupled symmetrically to the Cu$^{2+}$ ions carrying the relevant electronic spins $S=1/2$. As a result, the NMR relaxation rate $1/T_1$ filters out $q=\pm\pi$ components but conserves nonetheless the long-wavelength modes $q=0$ holding hydrodynamics. The applied field is directly proportional to the NMR frequency $\omega_0$ as per the Zeeman splitting.}
    \label{fig:nmr_exp_data}
\end{figure}

We revisit in Fig.~\ref{fig:nmr_exp_data} the experimental data of Fig.~3(d) in Ref.~\onlinecite{PhysRevLett.87.247202}. In this work, a power-law behavior of the form $1/T_1\propto H^{-\alpha}$ assuming $\alpha=0.5$ (corresponding to diffusion) was reported for the nearly ideal spin-$1/2$ Heisenberg antiferromagnets Sr$_2$CuO$_3$. Here, in addition to the diffusive behavior, we show the best superdiffusive fit of the form $\propto H^{-1/3}$, which is the expected behavior for the quantum spin-$1/2$ Heisenberg chain, based on today's knowledge. We also show the best constant fit of the form $\propto H^{0}$ corresponding to ballistic transport.

From a purely theoretical perspective, we expect ballistic spin transport in the infinite time limit due to the external magnetic field. However, the magnetic field being extremely small ($14$~T) compared to the spin exchange coupling in this compound ($J\simeq 2200$~K), the crossover might happen beyond the timescale related to the NMR frequency, making the dynamics look effectively supper-diffusive. The effect of the magnetic field needs to be precisely studied and we leave that for future work. For instance, for the low-energy physics studied in Refs.~\onlinecite{PhysRevLett.87.247202} and~\onlinecite{takigawa1997}, the effect of the magnetic field was irrelevant.

In any case, three data points are not enough to unambiguously identify the correct behavior, calling for new and dedicated NMR experiments on the issue of anomalous spin transport in one-dimensional spin chains. In particular, we believe that the present numerical abilities to efficiently simulate the microscopic dynamics of interacting 1D quantum models could greatly help in guiding experiments.

\subsection{The behavior $\Upsilon(T)\propto T^2$ for $T\ll J$ is compatible with experimental observations}

We approximate the real part of the spin-spin correlation $\mathfrak{Re}\,C(T,x=0,t)$ by $\Upsilon(T)t^{-2/3}$, which is the correct behavior in the long-time limit, see Fig.~2 in the main text. We get for the NMR relaxation rate,
\begin{equation}
    \frac{1}{T_1}\sim\int_0^{1/\omega_0}\mathfrak{Re}\,C(T,x=0,t)\,\mathrm{d}t~\sim~\Upsilon(T)\omega_0^{-1/3}~\Longrightarrow~\frac{1}{T_1}\sim T^2\omega_0^{-1/3}~~\mathrm{for}~~T\ll J,
    \label{eq:T1_longtime}
\end{equation}
where we found that $\Upsilon(T)\sim T^2$ for $T\ll J$, see Fig.~3(b) in the main text. As discussed in the main text, $\Upsilon(T)$ relates to the temperature dependence of the parameters of the KPZ equation: $\chi_\mathrm{s}\lambda_\mathrm{KPZ}^{-2/3}f_\mathrm{KPZ}(0)\sim\Upsilon(T)$.\\

In Fig.~4(a) of Ref.~\onlinecite{PhysRevLett.87.247202}, the authors find that for $T\ll J$, the NMR relaxation rate of Sr$_2$CuO$_3$ at fixed frequency $\omega_0$ may be approximated by an empirical form $1/T_1\approx aT+bT^2$ for $T\ll J$ with $a$ and $b$ fitting constants. Up to the term with linear temperature dependence $aT$, this is the behavior obtained in Eq.~\eqref{eq:T1_longtime}.

Neglecting the experimental data points for very low temperatures ($T\lesssim 100$ K), the experimental data of Fig.~4(a) in Ref.~\onlinecite{PhysRevLett.87.247202} is compatible with $1/T_1\sim T^2$. Substituting the real part of the correlator by its asymptotic behavior in Eq.~\eqref{eq:T1_longtime} becomes less and less valid at very low temperatures: the low-temperature physics of the real part of the correlator, not taken into account in the approximation of Eq.~\eqref{eq:T1_longtime} becomes dominant over high-temperature superdiffusive regime. In other words, in the time window $t\in[0,1/\omega_0]$, the two regimes coexist with the low-temperature one for $t\lesssim t^\star$ and the high-temperature one for $t\gtrsim t^\star$, with $t^\star\sim 1/T$ (see main text). In Eq.~\eqref{eq:T1_longtime}, it is assumed that the high-temperature regime is dominant. In this picture, we interpret the small flattening observed for very low temperatures ($T\lesssim 100$ K) in Fig.~4(a) of Ref.~\onlinecite{PhysRevLett.87.247202}, and which gives rise to the linear term $aT$, as the onset of low-temperature physics characterized by $1/T_1\simeq\ln^{-1/2}\bigl(J/T\bigr)$~\cite{takigawa1997,barzykin2001,dupont2016}. In fact, this logarithmic divergence was reported in Ref.~\onlinecite{takigawa1997} for the same compound (Sr$_2$CuO$_3$) for temperatures $T/J\lesssim 0.05$, corresponding to $T\simeq 100$ K, i.e., the regime where a linear term $aT$ is necessary to fit the experimental $1/T_1$ data.

For these reasons, we believe that the behavior $1/T_1\sim T^2$ reported in Eq.~\eqref{eq:T1_longtime} is compatible with earlier experimental measurements on Sr$_2$CuO$_3$~\cite{PhysRevLett.87.247202}, and relates to the temperature dependence of the parameters of the KPZ equation.

\section{Bond dimension convergence of the numerical simulations}

\begin{figure}[!h]
    \includegraphics[width=1.0\columnwidth]{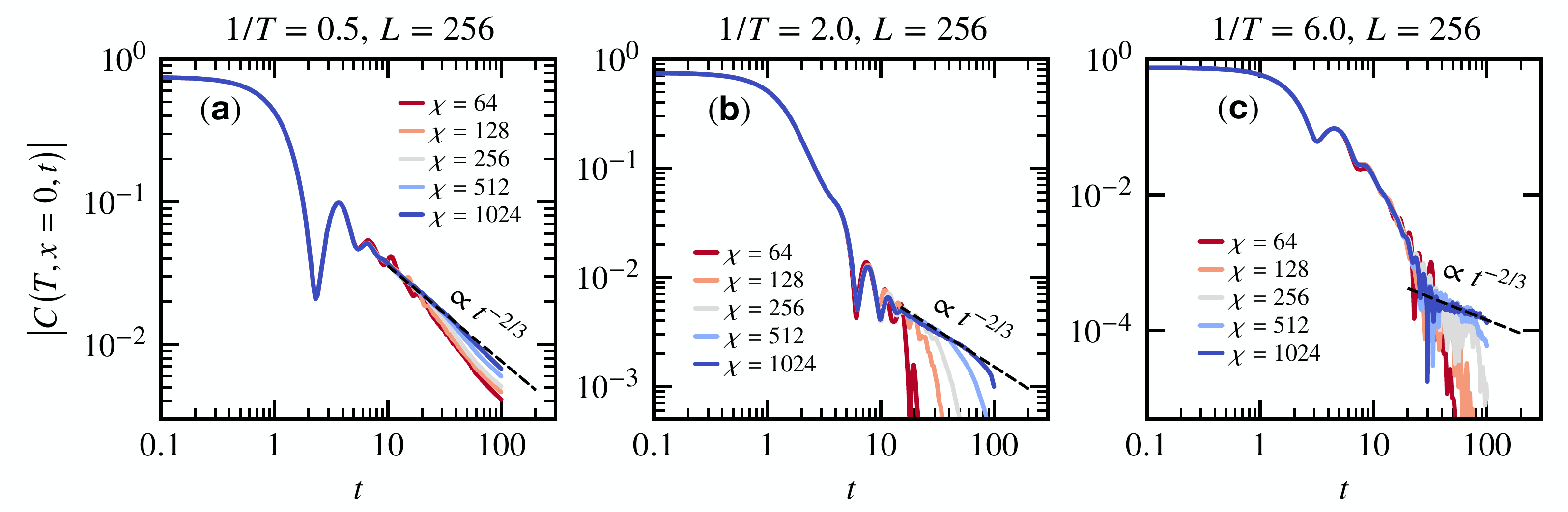} 
    \caption{Time dependence of the norm of the spin-spin correlation of Eq.~(2) in the main text at $x=0$ for various values of the bond dimension $\chi=64$, $128$, $256$, $512$, and $1024$. Simulations obtained for $L=256$ at three different temperatures (\textbf{a}) $1/T=0.5$, (\textbf{b}) $1/T=2.0$, and (\textbf{c}) $1/T=6.0$. At long time, it displays an algebraic decay $\propto t^{-2/3}$ (dashed black line).}
    \label{fig:bond_dimension}
\end{figure}

To understand the effect of the finite bond dimension $\chi$ on the numerical simulation, we performed the same calculations for $\chi=64$, $128$, $256$, $512$, and $1024$ (the larger, the better, and results in the main text correspond to $\chi=1024$). As one increases the bond dimension, the numerical data gets closer and closer to the expected $\propto t^{-2/3}$ power-law dependence at long-time, see Fig.~\ref{fig:bond_dimension}.

\section{System size convergence of the numerical simulations}

\begin{figure}[!h]
    \includegraphics[width=1.0\columnwidth]{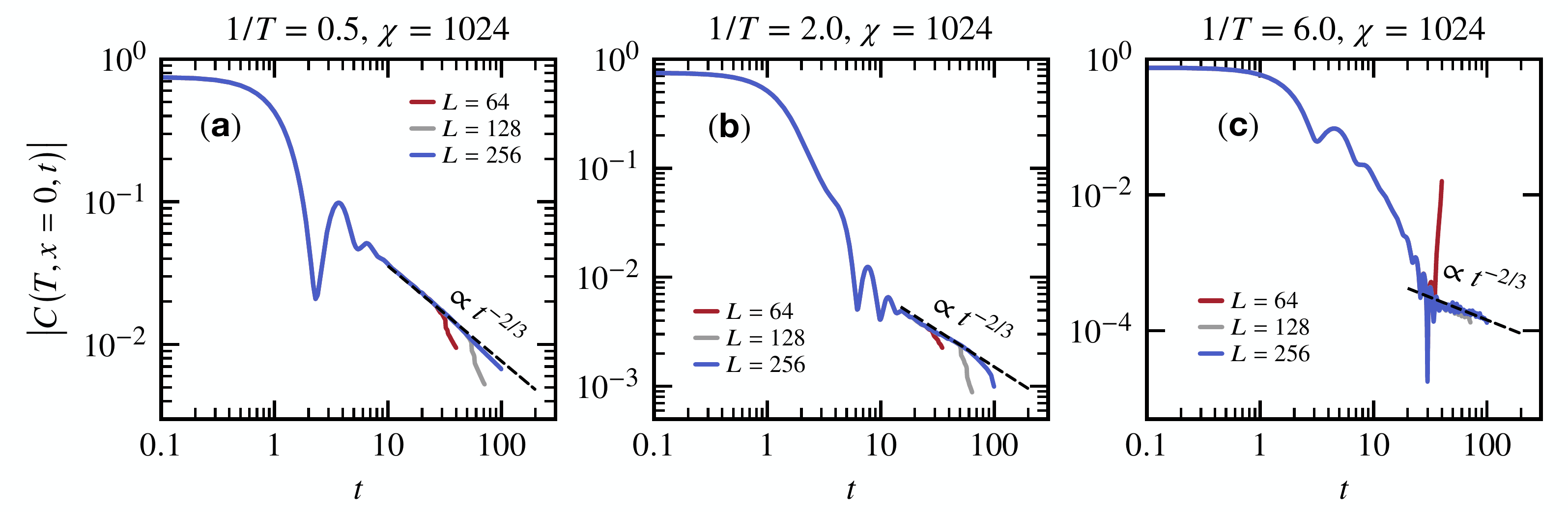} 
    \caption{Time dependence of the norm of the spin-spin correlation of Eq.~(2) in the main text at $x=0$ for various system sizes $L=64$, $128$, and $256$. Simulations obtained for a bond dimension $\chi=1024$ at three different temperatures (\textbf{a}) $1/T=0.5$, (\textbf{b}) $1/T=2.0$, and (\textbf{c}) $1/T=6.0$. At long time, it displays an algebraic decay $\propto t^{-2/3}$ (dashed black line).}
    \label{fig:system_size}
\end{figure}

By plotting data for increasing system sizes for the spin-spin correlation of Eq.~(2) in the main text at $x=0$, we see in Fig.~\ref{fig:system_size} that finite-size effects only take place at times $t\approx L/2$, reminiscent of the light-cone structure (data available for up to $L=256$). Data at $|x|>0$ in Fig.~4 of the main text are shown up to $|x|=25$, which is still far away from the system boundary at $|x|=128$. Therefore, the conclusions drawn in the manuscript are independent of the system size.

\section{Real part versus imaginary part of the dynamical spin-spin correlation}

\begin{figure}[!h]
    \includegraphics[width=1.0\columnwidth]{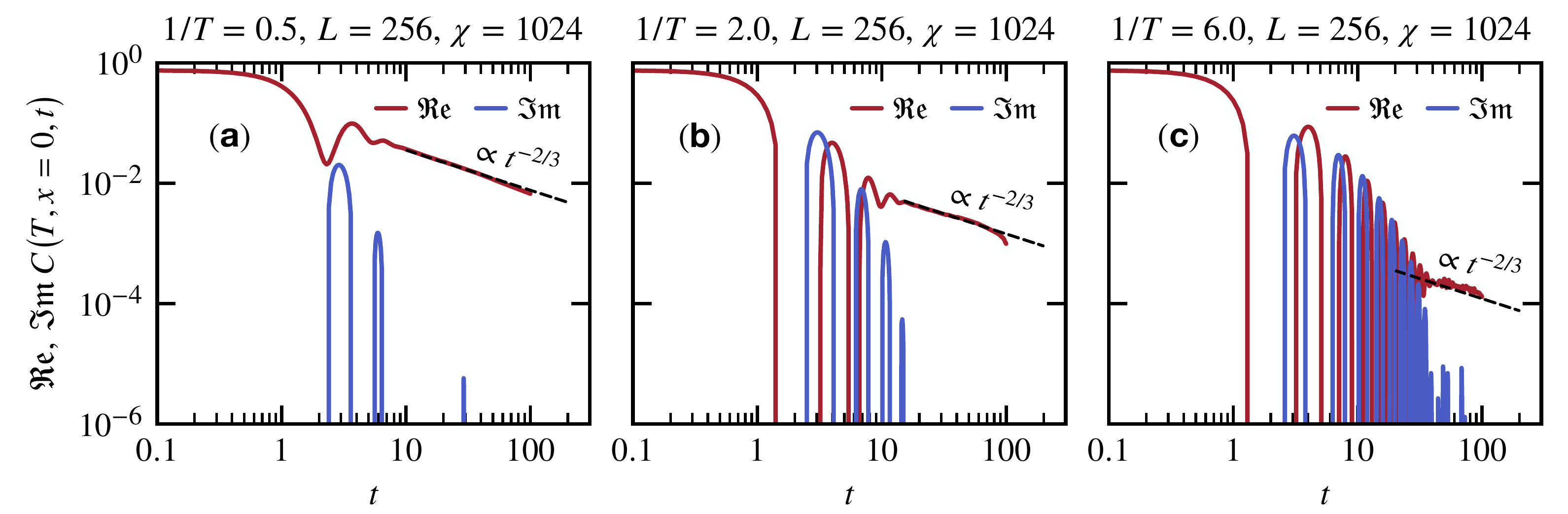} 
    \caption{Time dependence of the real part ``$\mathfrak{Re}$'' and imaginary part ``$\mathfrak{Im}$'' part of the spin-spin correlation of Eq.~(2) in the main text at $x=0$. Simulations obtained for $L=256$ and $\chi=1024$ at three different temperatures (\textbf{a}) $1/T=0.5$, (\textbf{b}) $1/T=2.0$, and (\textbf{c}) $1/T=6.0$. We observe that the superdiffusive power-law regime $\propto t^{-2/3}$ only holds for the real part (dashed black line) and that in this regime we have $\vert\mathfrak{Im}\,C(T,x=0,t)\vert\ll\vert\mathfrak{Re}\,C(T,x=0,t)\vert$.}
    \label{fig:real_imaginary_part}
\end{figure}

While we display the norm of the spin-spin correlation function in the main text, we compute both the real and imaginary parts. We show them independently in Fig.~\ref{fig:real_imaginary_part}. We observe that the real part hosts the characteristic power-law dependence $\propto t^{-2/3}$, not the imaginary part. In fact, in the hydrodynamics regime, we find that $\vert\mathfrak{Im}\,C(T,x,t)\vert\ll\vert\mathfrak{Re}\,C(T,x,t)\vert$, meaning that at long time, the imaginary part plays no role in the superdiffusive dynamics of the spin-$1/2$ Heisenberg chain.

\section{Low-temperature versus zero temperature}

\begin{figure}[!h]
    \includegraphics[width=1\columnwidth]{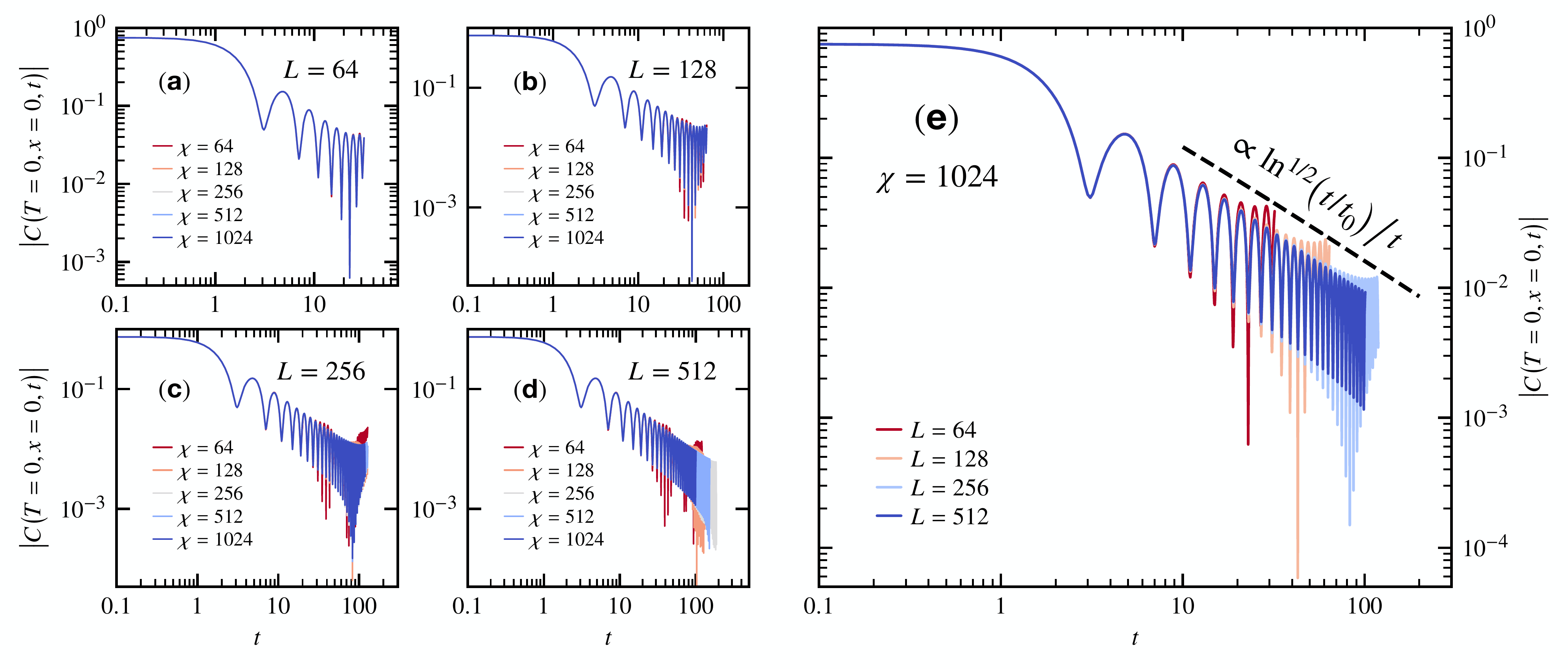} 
    \caption{Time dependence of the norm of the spin-spin correlation of Eq.~(2) in the main text at $x=0$ for various system sizes $L=64$, $128$, and $256$ and bond dimensions $\chi=64$, $128$, $256$, $512$, and $1024$ at zero temperature ($T=0$). (\textbf{a}) $L=64$, (\textbf{b}) $L=128$, (\textbf{c}) $L=256$, and (\textbf{d}) $L=512$ for various bond dimensions $\chi$. (\textbf{e}) $\chi=1024$ for various system sizes $L$. The dashed line is a fit of the form $\propto\ln^{1/2}\bigl(t/t_0\bigr)\bigr/t$, with $t_0\approx 0.5$ a fitting parameter.}
    \label{fig:zero_temperature}
\end{figure}

By plotting the different system sizes for the spin-spin correlation of Eq.~(2) in the main text at $x=0$ and $T=0$, see Fig.~\ref{fig:zero_temperature}(e) we show that the ``flattening'' observed at long times is a finite size effect. In Figs.~\ref{fig:zero_temperature}(a)--\ref{fig:zero_temperature}(d) we show the effect of the finite bond dimension, which is qualitatively very small.

Our data confirm the $\propto 1/t$ decay (up to logarithmic corrections) of the $x=0$ spin-spin correlation at zero temperature for the spin-$1/2$ Heisenberg chain~\cite{pereira2008,pereira2012} in Fig.~\ref{fig:zero_temperature}(e). Including logarithmic corrections, the decay follows $\propto\ln^{1/2}\bigl(t/t_0\bigr)\bigr/t$, with $t_0\approx 0.5$ a fitting parameter.\\

\begin{figure}[t]
    \includegraphics[width=0.6\columnwidth]{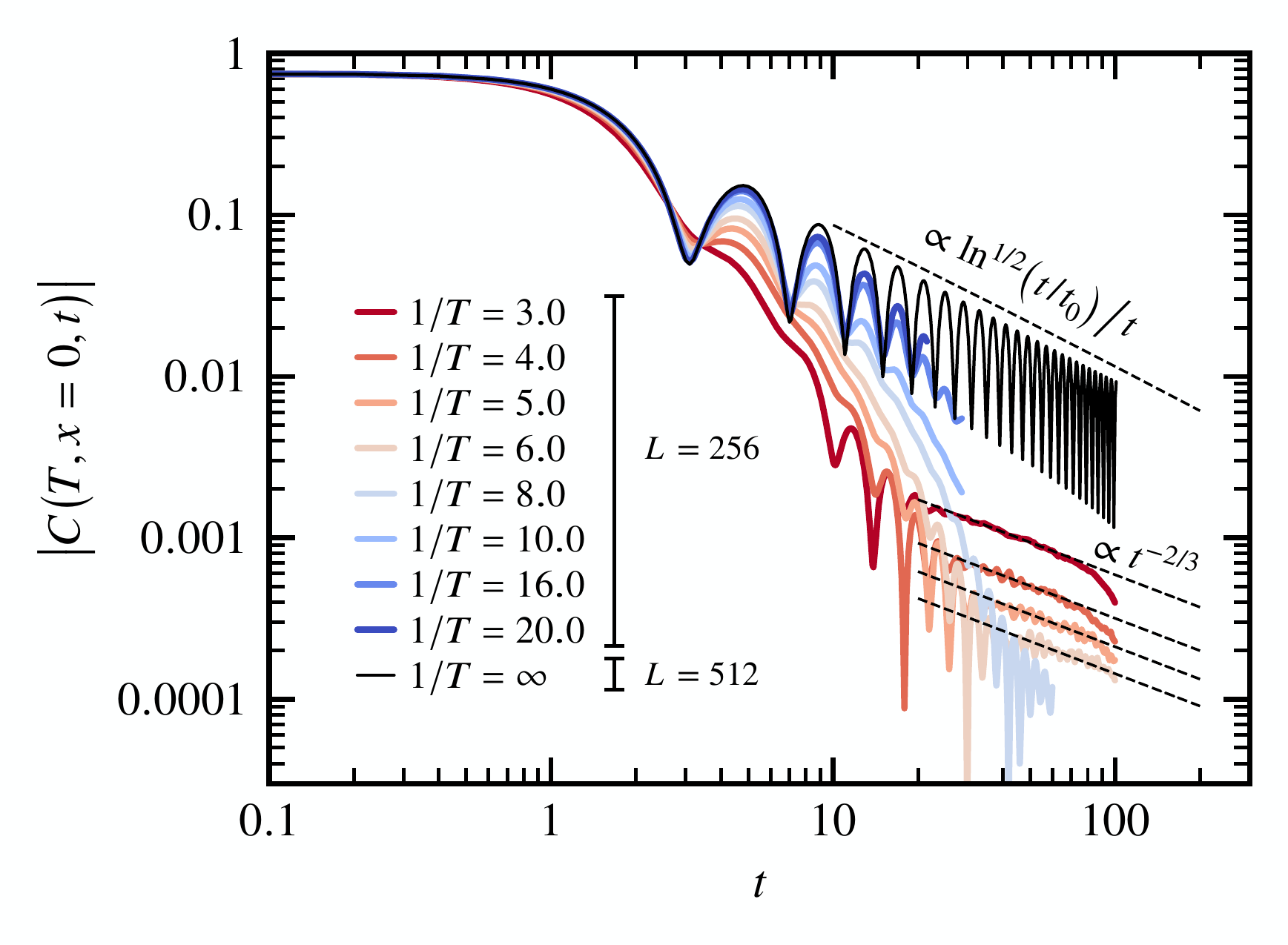} 
    \caption{Time dependence of the norm of the spin-spin correlation of Eq.~(2) in the main text at $x=0$ for various temperatures $T$. Simulations obtained for $L=256$ with $\chi=1024$ at finite temperature and for $L=512$ with $\chi=1024$ at zero temperature. Same data as in Fig.~2 of the main text plus the zero temperature ($1/T=\infty$), $1/T=8.0$, $1/T=10.0$, $1/T=16.0$, and , $1/T=20.0$ data. The dashed line next to the zero temperature data is a fit of the form $\propto\ln^{1/2}\bigl(t/t_0\bigr)\bigr/t$, with $t_0\approx 0.5$ a fitting parameter.}
    \label{fig:autocorrelation_lowT}
\end{figure}

We also confirm that the finite-temperature data (down to $1/T=6.0$ in the main text) is actually not small enough to observe the genuine low-temperature dynamics. We see in Fig.~\ref{fig:autocorrelation_lowT} that at least $1/T\gtrsim 20.0$ is required to have an overlap between zero-temperature and finite-temperature data in a reasonable time window. This rather slow convergence of the finite-temperature data onto the zero-temperature ones is also observed for the spatial dependence at $t=0$ in Fig.~\ref{fig:correlation_length}(a). It is understood from the absolute value of the thermal correlation length of Eq.~\eqref{eq:correlation_length}.\\

Note that the numerical simulations for zero temperature $T=0$ are carried out with a slightly different method than for $T>0$. In particular, we do not need to use the trick representing a mixed state as a pure state in an enlarged Hilbert space, the state at $T=0$ being a pure state (it is the ground state). The ground state is obtained with the density matrix renormalization group algorithm~\cite{PhysRevLett.69.2863,schollwock2011,itensor}, and the time evolution is then performed using time-evolving block decimation algorithm~\cite{PhysRevLett.93.040502} along with a fourth-order Trotter decomposition~\cite{hatano2005} with step $\delta=0.1$ leading to a negligible discretization error $O\bigl(\delta^5\bigr)$.

\newpage

\section{Temperature dependence of the correlation length}

\begin{figure}[!h]
    \includegraphics[width=0.7\columnwidth]{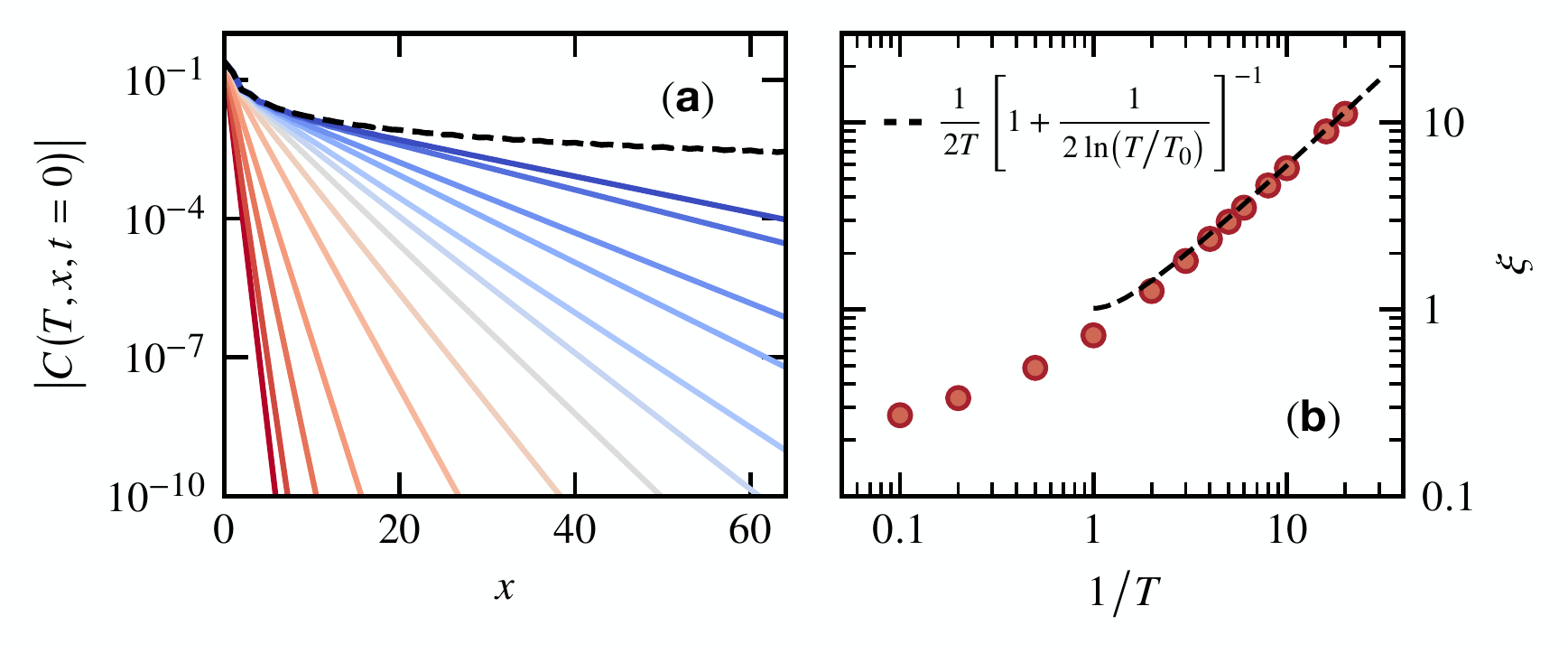} 
    \caption{(\textbf{a}) Spatial dependence of the norm of the spin-spin correlation of Eq.~(2) in the main text at $t=0$ for system size $L=256$ and bond dimension $\chi=1024$. From the lower left corner to the upper right one, the solid lines correspond to the following temperatures: $1/T=0.1$, $0.2$, $0.5$, $1.0$, $2.0$, $3.0$, $4.0$, $5.0$, $8.0$, $10.0$, $16.0$, and $20.0$. The dashed line is the zero temperature data ($1/T=\infty$). Except for the zero temperature data which decay as $\propto\ln^{1/2}\bigl(x/x_0\bigr)\bigr/x$, the finite temperature data decay exponentially at long distance $x$. A fit of the form $\propto\exp\bigl(-x/\xi\bigr)$ gives access to the correlation length $\xi$. (\textbf{b}) Correlation length $\xi$ plotted versus the inverse temperature $1/T$. The dashed line is the expression of Eq.~\eqref{eq:correlation_length} valid as $T\to 0$ and derived in Ref.~\onlinecite{nomura1991} with $T_0\approx 2.68$.}
    \label{fig:correlation_length}
\end{figure}

The thermal correlation length $\xi$ of the spin-$1/2$ Heisenberg chain diverges at low temperature as $1/T$, plus additional log corrections which at first order gives~\cite{nomura1991},
\begin{equation}
    \xi\simeq\frac{1}{2T}\left[1+\frac{1}{2\ln\bigl(T\bigr/T_0\bigr)}\right]^{-1},
    \label{eq:correlation_length}
\end{equation}
with $T_0\approx 2.68$ a nonuniversal constant. Here, we have used the value of $T_0$ obtained in Ref.~\onlinecite{nomura1991} computed by the thermal Bethe ansatz. The agreement in Fig.~\ref{fig:correlation_length} is extremely good for $1/T\gtrsim 5$. Yet, for the range of temperatures considered in this work, the data could be fitted equally well without the log corrections.

The data of Fig.~\ref{fig:correlation_length} together with Eq.~\eqref{eq:correlation_length} shows why one needs to go to extremely low temperatures to observe the genuine low-temperature physics of the spin-$1/2$ Heisenberg chain: the prefactor of the temperature dependence of the correlation length $\xi\simeq 1/2T$ is small. It explains why in the time-dependent data of Fig.~\ref{fig:autocorrelation_lowT} it is difficult to observe a good overlap between zero-temperature and finite-temperature data for the temperatures accessible in this work; this overlap is equally hard to observe for the spatial dependence in Fig.~\ref{fig:correlation_length}(a).

\section{Temperature dependence of the parameters of the KPZ equation}

\begin{figure}[!h]
    \includegraphics[width=1.0\columnwidth]{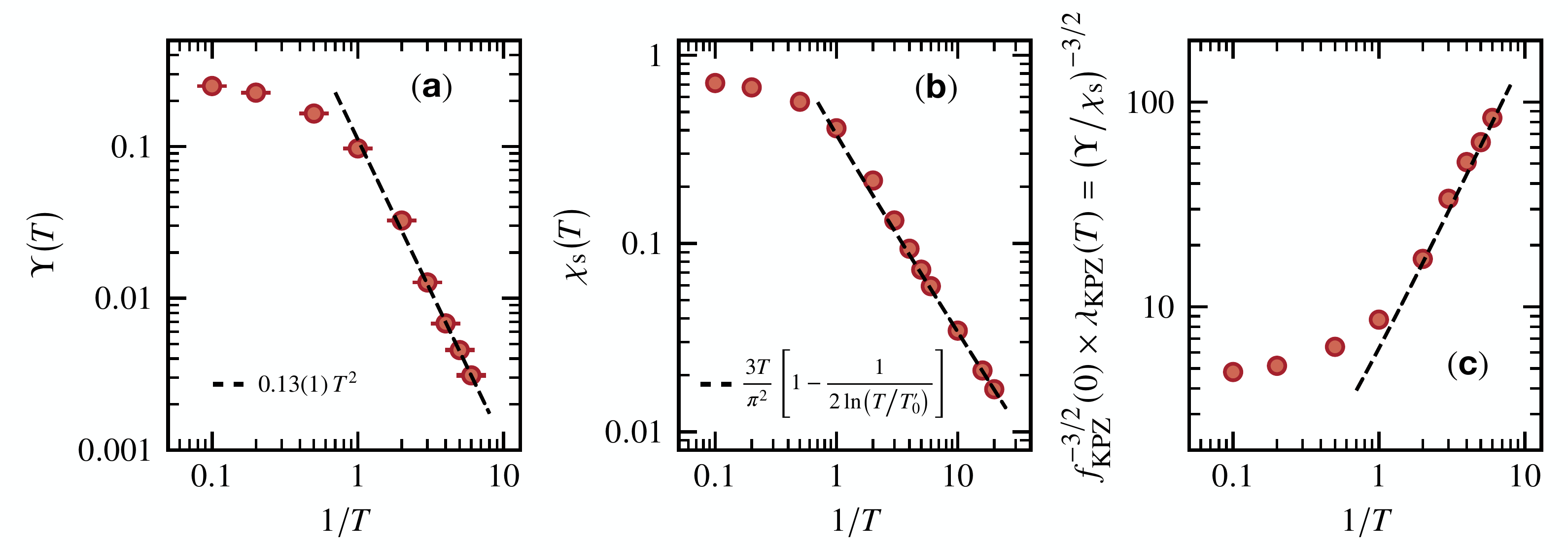} 
    \caption{(\textbf{a}) Same as Fig.~3(a) in the main text. Temperature dependence of the prefactor $\Upsilon(T)$ of the algebraic decay $\propto t^{-2/3}$ for superdiffusive hydrodynamics at $x=0$. At low temperatures $T\lesssim 1$, it follows a quadratic dependence $\propto T^2$ (dashed line). (\textbf{b}) Temperature dependence of the static spin susceptibility $\chi_\mathrm{s}(T)$ defined in Eq.~\eqref{eq:spin_susc}. The dashed line is the expression reported in Eq.~\eqref{eq:spin_susc} valid as $T\to 0$ and derived in Ref.~\onlinecite{eggert1994} with $T_0'\approx 7.7$. (\textbf{c}) By identifying $\Upsilon(T)=\chi_\mathrm{s}\lambda_\mathrm{KPZ}^{-2/3}f_\mathrm{KPZ}(0)$, we get $f^{-3/2}_\mathrm{KPZ}(0)\times\lambda_\mathrm{KPZ}(T)=(\Upsilon/\chi_\mathrm{s})^{-3/2}$, and plot its temperature dependence. Based on the reported results for $\Upsilon(T)$ and $\chi_\mathrm{s}(T)$, the dashed line has a dominant $\propto T^{-3/2}$ behavior plus additional log corrections originating from $\chi_\mathrm{s}(T)$, see Eq.~\eqref{eq:lambda_kpz}.}
    \label{fig:kpz_parameters}
\end{figure}

By identifying the prefactor of $C_\mathrm{KPZ}(x=0,t)$ of Eq.~(4) in the main text with the prefactor $\Upsilon(T)$ of the power-law decay $\propto t^{-2/3}$ shown in Fig.~\ref{fig:kpz_parameters}(a), we find $\Upsilon(T)=0.13(1)T^2$ for $T\lesssim 1$, and therefore that $\chi_\mathrm{s}\lambda_\mathrm{KPZ}^{-2/3}f_\mathrm{KPZ}(0)\propto T^2$. It is established that in this temperature range, the static spin susceptibility of the spin-$1/2$ Heisenberg chain takes the form~\cite{eggert1994},
\begin{equation}
    \chi_\mathrm{s}\bigl(T\bigr)=\sum\nolimits_x\Bigl\langle\hat{\boldsymbol{S}}_x\cdot\hat{\boldsymbol{S}}_0\Bigr\rangle\simeq\frac{3T}{\pi^2}\left[1-\frac{1}{2\ln\bigl(T\bigr/T_0'\bigr)}\right],
    \label{eq:spin_susc}
\end{equation}
with $T_0'\approx 7.7$ a nonuniversal constant obtained in Ref.~\onlinecite{eggert1994} through Bethe ansatz (there are also higher order log corrections). The form of Eq.~\eqref{eq:spin_susc} is verified in Fig.~\ref{fig:kpz_parameters}(b). Isolating $\lambda_\mathrm{KPZ}$, we find that,
\begin{equation}
    \lambda_\mathrm{KPZ}\bigl(T\bigr)=\left(\frac{\Upsilon}{\chi_\mathrm{s}f_\mathrm{KPZ}(0)}\right)^{-3/2}\simeq\left\{\frac{0.13(1)T\pi^2}{3f_\mathrm{KPZ}(0)}\left[1-\frac{1}{2\ln\bigl(T\bigr/T_0'\bigr)}\right]^{-1}\right\}^{-3/2},
    \label{eq:lambda_kpz}
\end{equation}
which we plot in Fig.~\ref{fig:kpz_parameters}(c). The high-temperature data points are compatible with Ref.~\onlinecite{denardis2020b}.

\section{Spatial dependence of the crossover time}

\begin{figure}[!h]
    \includegraphics[width=0.45\columnwidth]{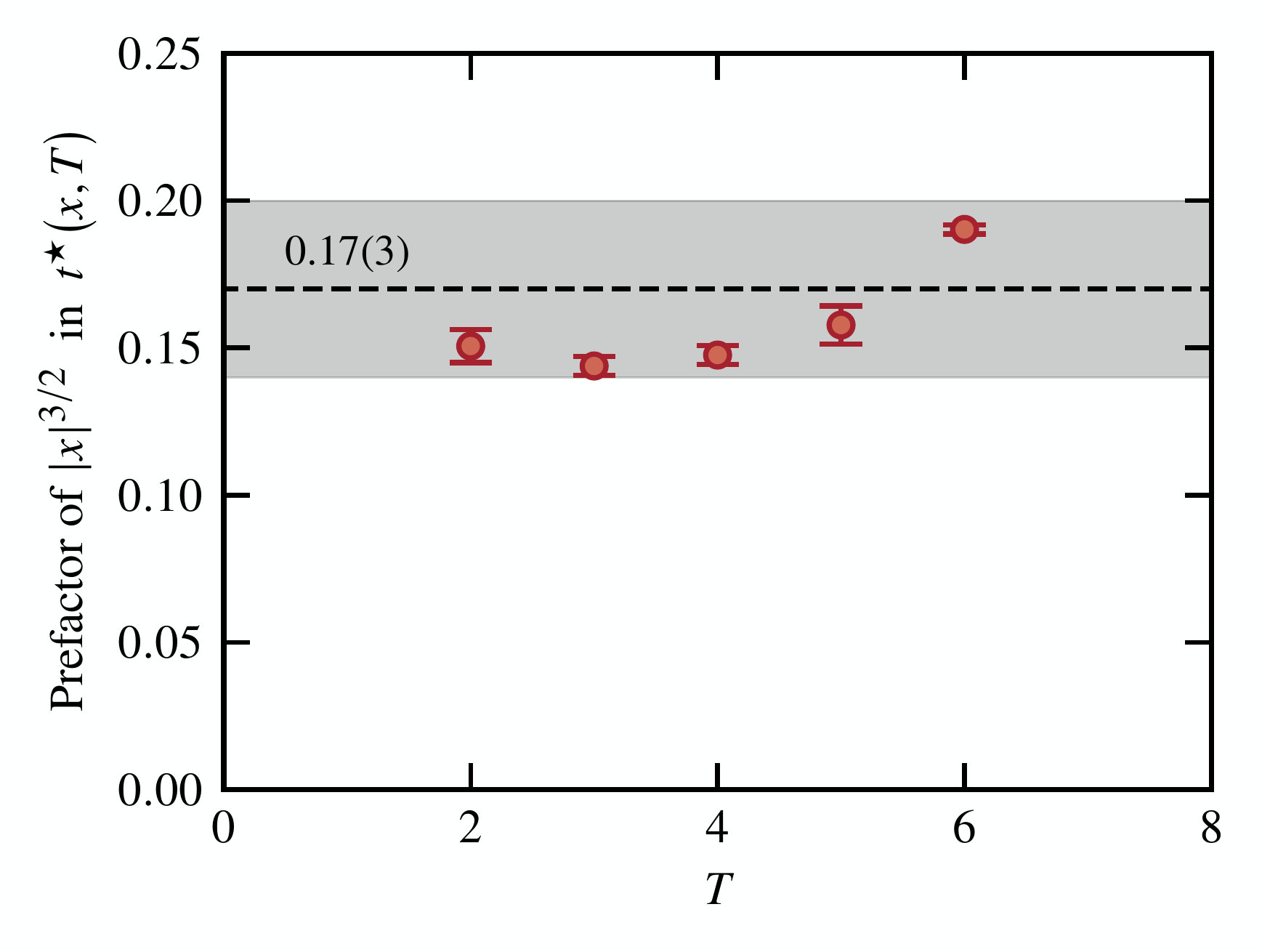} 
    \caption{Temperature dependence of the prefactor of the spatial dependence of $t^\star\bigl(x,T\bigr)$ of Eq.~(5) in the main text. The prefactor, which is plotted here, is extracted from a least-square fitting of the data of Fig.~4(b) of the main text. It is roughly independent of temperature with a value compatible with $0.17(3)$.}
    \label{fig:prefactor_temperature}
\end{figure}

We show in Fig.~\ref{fig:prefactor_temperature} the temperature dependence of the prefactor of the spatial dependence of $t^\star\bigl(x,T\bigr)$ of Eq.~(5) in the main text, i.e., the term $\propto\vert{x}\vert^{3/2}$. It is extracted from a least-square fitting of the data of Fig.~4(b) of the main text. While it fluctuates slightly from one temperature to the next, there is no clear trend observed, and the data is consistent with a constant prefactor with value $0.17(3)$.

\section{Extraction of the crossover time from the microscopic simulations}

There is a degree of appreciation in defining the crossover time $t^\star\bigl(T,x\bigr)$ from the numerical simulations, precisely because it is related to a crossover and not, e.g., a sharp transition. We have defined two quantities to extract $t^\star\bigl(T,x\bigr)$,
\begin{equation}
    \mathsf{R}_\mathsf{A}\bigl(T,x,t\bigr)=\overline{\left[\frac{\Bigl\vert\bigl\vert C\bigl(T,x,t\bigr)\bigr\vert-\Upsilon\bigl(T,x\bigr)t^{-3/2}\Bigr\vert}{\bigl\vert C\bigl(T,x,t\bigr)\bigr\vert}\right]}^{t\pm 2}\quad\mathrm{,~and}\quad\mathsf{R}_\mathsf{B}\bigl(T,x,t\bigr)=\overline{\left\vert\frac{\mathfrak{Im}\;C\bigl(T,x,t\bigr)}{\mathfrak{Re}\;C\bigl(T,x,t\bigr)}\right\vert}^{t\pm 2}.
    \label{eq:extracting_tstar}
\end{equation}
The first one returns the relative difference between the norm of the spin-spin correlation of Eq.~(2) in the main text with the superdiffusive decay $\Upsilon\bigl(T,x\bigr)t^{-3/2}$. The second one returns the relative weight of the imaginary part $\mathfrak{Im}$ versus the real part $\mathfrak{Re}$ of the spin-spin correlation of Eq.~(2) in the main text. In Eq.~\eqref{eq:extracting_tstar}, $T$ and $x$ are set to given values and the quantities are looked at versus the time $t$. The symbol $\overline{(-)}^{t\pm 2}$ means that the data at time $t$ actually corresponds an average from the range $\in[t-2,t+2]$. The effect is to smoothen the local oscillations in $C\bigl(T,x,t\bigr)$, and make the extraction more reliable.

\begin{figure}[!h]
    \includegraphics[width=0.8\columnwidth]{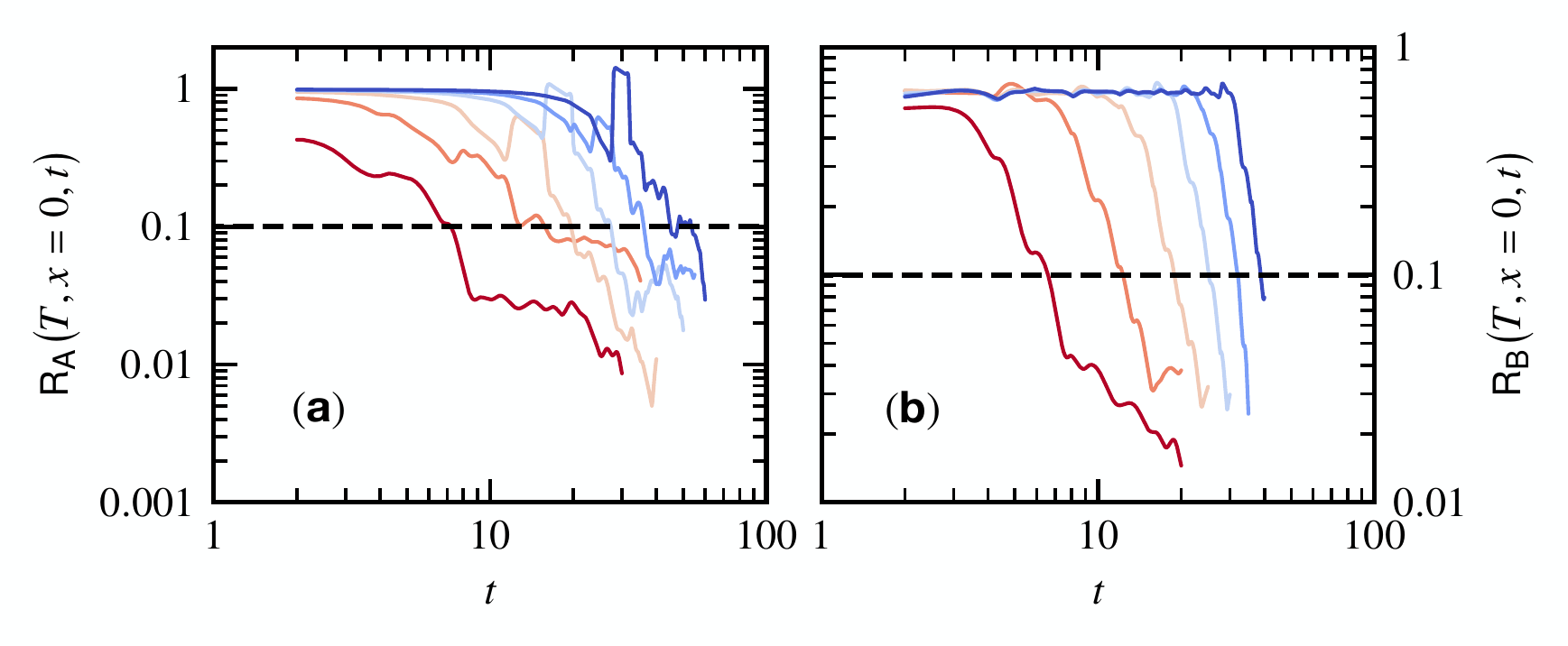} 
    \caption{Panels (\textbf{a}) and (\textbf{b}) correspond to the quantities $\mathsf{R}_\mathsf{A}\bigl(T,x=0,t\bigr)$ and $\mathsf{R}_\mathsf{B}\bigl(T,x=0,t\bigr)$ of Eq.~\eqref{eq:extracting_tstar}, respectively. The system size is $L=256$ and the bond dimension $\chi=1024$. From the lower left corner to the upper right one, the solid lines correspond to the following temperatures: $1/T=1.0$, $2.0$, $3.0$, $4.0$, $5.0$, and $6.0$. The intersection of the data with the horizontal dashed line at $\mathsf{R}_\mathsf{A,B}\bigl(T,x=0,t\bigr)=0.1$ is used to extract the crossover time $t^\star\bigl(T,x\bigr)$.}
    \label{fig:extracting_tstar}
\end{figure}

The time at which $\mathsf{R}_\mathsf{A}\bigl(T,x,t\bigr)$ and $\mathsf{R}_\mathsf{B}\bigl(T,x,t\bigr)$ hit the value $0.1$ is used as the definition of the crossover time $t^\star\bigl(T,x\bigr)$. The definition of $\mathsf{R}_\mathsf{B}\bigl(T,x,t\bigr)$ uses the fact that in the hydrodynamics regime we have $\vert\mathfrak{Im}\,C(T,x,t)\vert\ll\vert\mathfrak{Re}\,C(T,x,t)\vert$. Both quantities lead to comparable estimates of $t^\star\bigl(T,x\bigr)$. The error bar reported on $t^\star\bigl(T,x\bigr)$ reflects the small difference between the two estimates. As an example, we show $\mathsf{R}_\mathsf{A}\bigl(T,x,t\bigr)$ and $\mathsf{R}_\mathsf{B}\bigl(T,x,t\bigr)$ for $x=0$ in Fig.~\ref{fig:extracting_tstar}. The corresponding value $t^\star\bigl(T,x=0\bigr)$ is reported in Fig.~3(a) in the main text.

\end{document}